\documentclass[USenglish,twocolumn]{article}
\usepackage[utf8]{inputenc}
\usepackage[big,online]{dgruyter}
\usepackage[sort&compress,square,numbers]{natbib}
\usepackage{times}

\usepackage{xcolor}

\begin{document}

  \DOI{10.1515/}
  \openaccess
  \pagenumbering{gobble}

\title{Automatic Tissue Differentiation in Parotidectomy using Hyperspectral Imaging}
\runningtitle{Automatic Tissue Differentiation using Hyperspectral Imaging}

\author*[1]{Eric L.~Wisotzky}
\author[2]{Alexander Schill}
\author[2]{Anna Hilsmann} 
\author[3]{Peter Eisert}
\author[4]{Michael Knoke}
\runningauthor{E.~Wisotzky et al.}

\affil[1]{\protect\raggedright 
  Fraunhofer Heinrich-Hertz-Institut HHI, Humboldt Universit\"at \&  Charit\'e -- Universit\"atsmedizin Berlin, Germany, e-mail: \{name\}.\{surname\}@hhi.fraunhofer.de}
\affil[2]{\protect\raggedright
  Computer Vision \& Graphics, Fraunhofer Heinrich-Hertz-Institut HHI, Berlin, Germany}
\affil[3]{\protect\raggedright
  Fraunhofer Heinrich-Hertz-Institut HHI \& Humboldt Universit\"at zu Berlin, Germany}
\affil[4]{\protect\raggedright 
  Klinik für Hals-, Nasen-, Ohrenheilkunde, Charit\'e -- Universit\"atsmedizin Berlin, Germany;
\textbf{P.E.} and \textbf{M.K.} contributed equally to this work.}

\abstract{
In head and neck surgery, continuous intraoperative tissue differentiation is of great importance to avoid injury to sensitive structures such as nerves and vessels. Hyperspectral imaging (HSI) with neural network analysis could support the surgeon in tissue differentiation.
A 3D Convolutional Neural Network with hyperspectral data in the range of $400-1000$ nm is used in this work. The acquisition system consisted of two multispectral snapshot cameras creating a stereo-HSI-system. For the analysis, 27 images with annotations of glandular tissue, nerve, muscle, skin and vein in 18 patients undergoing parotidectomy are included. Three patients are removed for evaluation following the leave-one-subject-out principle. The remaining images are used for training, with the data randomly divided into a training group and a validation group. 
In the validation, an overall accuracy of $98.7\%$ is achieved, indicating robust training. In the evaluation on the excluded patients, an overall accuracy of $83.4\%$ has been achieved showing good detection and identification abilities.
The results clearly show that it is possible to achieve robust intraoperative tissue differentiation using hyperspectral imaging. Especially the high sensitivity in parotid or nerve tissue is of clinical importance. It is interesting to note that vein was often confused with muscle. This requires further analysis and shows that a very good and comprehensive data basis is essential. This is a major challenge, especially in surgery.
}

\keywords{hyperspectral imaging, multispectral imaging, tissue differentiation, 3D-CNN, tissue classification.}

\maketitle

\section{Introduction} 

Parotidectomy, a surgical procedure primarily performed to address tumors or pathological conditions affecting the parotid gland, presents surgeons with a unique set of challenges and complexities. The parotid gland, located near the ear, is intricately connected to vital structures such as facial nerves and blood vessels. One of the foremost challenges in parotidectomy is the proximity of the facial nerve, which traverses through the gland and branches out into numerous smaller nerves responsible for facial expression. Accidental damage to these nerves during surgery can result in significant morbidity, including facial weakness or paralysis. Moreover, the parotid gland's anatomical variability and irregular shape further complicate the surgical procedure, making precise tumor resection while preserving surrounding vital structures a delicate balancing act. Additionally, distinguishing between tumor tissue and healthy glandular tissue can be challenging due to their similar appearance, heightening the risk of incomplete tumor removal or unnecessary tissue excision. 

These challenges underscore the critical need for intraoperative guidance tools enabling the surgeon to precisely and objectively discriminate between different kinds of soft tissue. Iatrogenic injuries to structures located in the surgical area, such as nerves and vessels could be prevented using new imaging tools such as hyperspectral imaging \cite{ClancySSI}. 

Hyperspectral imaging (HSI) is an optical imaging technology showing tremendous potential across various fields, including medicine and surgery \cite{LuHSIreview}. Similar to conventional RGB imaging, it is non-invasive but allows to acquire a multitude of spectral bands across the electromagnetic spectrum \cite{ShapeyReview}. This results in rich, detailed data that can reveal subtle differences in tissue composition, aiding in the identification of abnormalities and anomalies.

In the realm of surgery, HSI offers a transformative approach to intraoperative tissue visualization \cite{FischerOrgan}. By leveraging the unique spectral signatures of different tissues, HSI enables surgeons to discern between different healthy and diseased tissues \cite{WisotzkySensor}. However, HSI produces a high amount of data, making it hard for a human to analyze directly without help.
By analyzing the distinct spectral characteristics of, e.g., nerves using neural networks (NN), HSI could assist surgeons in accurately identifying nerve pathways amidst surrounding tissues during parotidectomy. This capability has the potential to significantly reduce the risk of nerve damage and improve surgical outcomes for patients undergoing parotidectomy procedures.

The aim of this work is to investigate the utility of hyperspectral imaging in combination with a convolutional neural network (CNN) for tissue identification during parotidectomy. This paper will introduce the HSI setup and the used 3D NN structure in the next chapter. Following, we present the collected data of nerves, glandular tissue, vessels, fat, and skin. In chapter \ref{sec:res}, we discuss our results, finalized by a conclusion in chapter \ref{sec:conc}.

\section{Methods}
\subsection{Stereo-HSI Setup}
In order to achieve a hyperspectral image with high spatial resolution and real-time capabilities, we are using multispectral snapshot cameras (Ximea GmbH, Germany) \cite{EbnerSnapshot}. In order to achieve hyperspectral information of the scene, we combine two multispectral snapshot mosaic cameras with $16$ and $25$ spectral bands, covering different spectral intervals ($400-650$ nm and $475-975$ nm, respectively \cite{WisotzkyValidation,MuhleComparison}) into a stereo setup. To fuse the captured data of both cameras into one HSI data cube, we first demosaic the two snapshot mosaic images individually \cite{WisotzkyDemosaicking, WisotzkyDemosaicking2}, then calculate a dense registration between the two stereo views and finally fuse the $16$ spectral bands of the left camera to the spectral data of the right camera based on the calculated warps \cite{WisotzkyStereoHSI, WisotzkyStereoHSI2}. Thus, a hyperspectral data cube results with 41 spectral bands and a spatial image resolution of about $2000\times1000$ pixels.
As illumination source for our system, the present surgical light and the diffuse ceiling light is used. The achieve reflectance behavior of the captured scene, a strict calibration pipeline is followed including white image correction before the capturing \cite{WisotzkyValidation}.

\subsection{Classification Network}
Convolutional neural network (CNN) showed great potential in image segmentation and data classification \cite{Ortac21}.
As our data show a three dimensional structure (3D) with two spatial and one spectral dimension, we used 3D convolutional layers in the network. The complete network consists of six 3D convolutional layers followed by three fully connected layers. The initial layer receives a patch of $31\times31\times41$ pixels (H$\times$W$\times\lambda$) as input, and its primary function is to extract relevant features crucial for classification tasks. These extracted features are then forwarded to the subsequent layers, where more abstract and high-level features are computed based on the input 3D features. This process iterates through the network until reaching the final layer. Ultimately, the features extracted at the last layer are utilized to make the classification decision between the five different tissue types nerve, gland, muscle, vein, and skin.

\subsection{Imaging and Data Annotation}
The data capturing is performed during 18 parotidectomies. After dissection of the entire situs and successful identification of the individual branches of the facial nerve, the different tissue types have been localized according to present anatomy. The focus here is placed on the predominantly present and most important types of tissue in parotidectomy: nerve, glandular tissue, vein, muscle, and skin.

\begin{figure}[t]
	\includegraphics[width=.855\columnwidth]{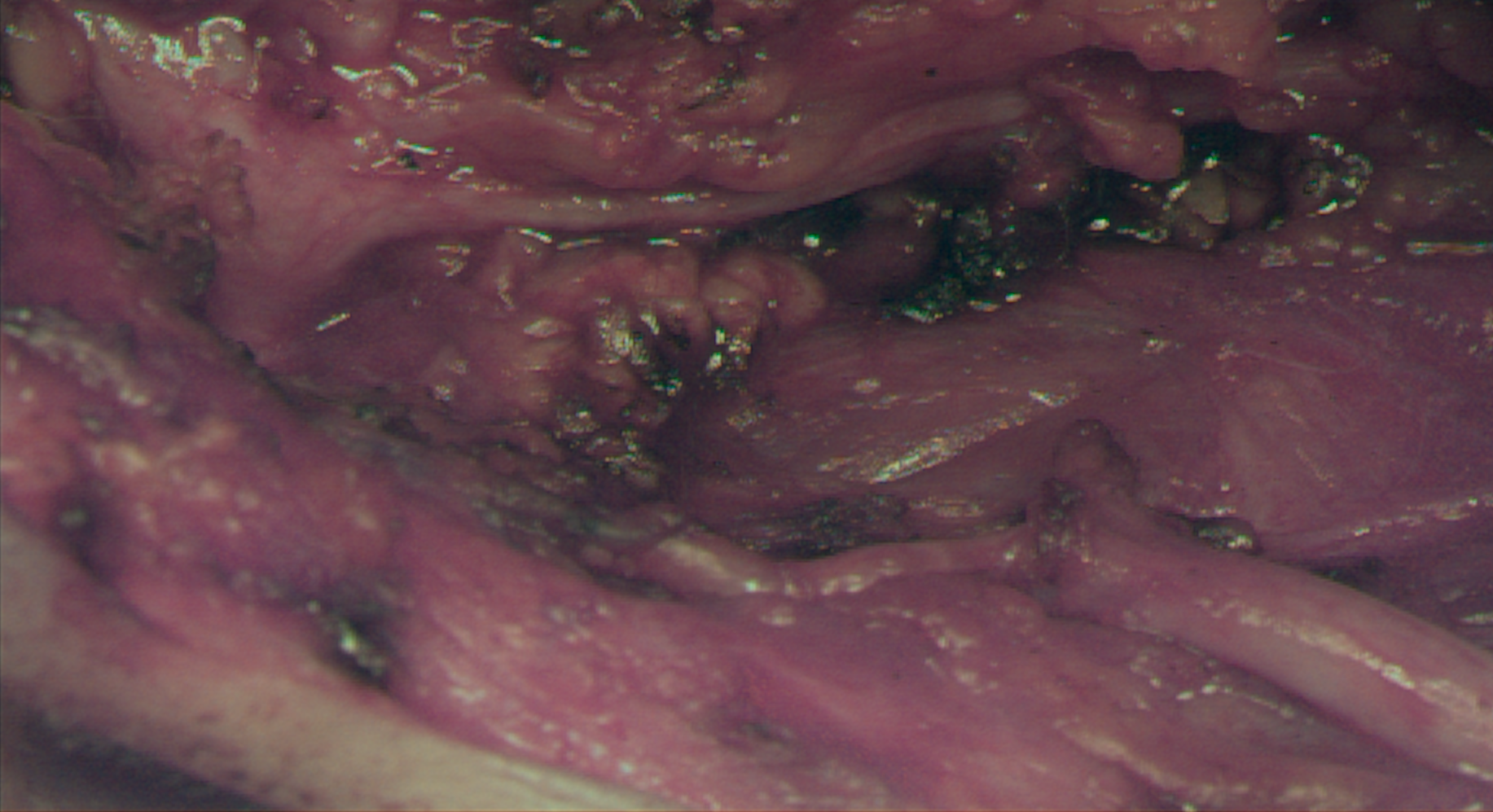}
	\caption{The classical RGB representation of an captured hyperspectral image.}
	\label{img:rgb}
\end{figure}

The system has be positioned in a way all wanted tissue types are present in the image. If that was not possible within one scene, several different images has been acquired of that patient to capture from every patient all aimed tissue types. Therefore, 27 images were captured in total. An RGB representation of a captured HSI cube is presented in Figure \ref{img:rgb}.

\begin{figure}[t]
	\includegraphics[width=\columnwidth]{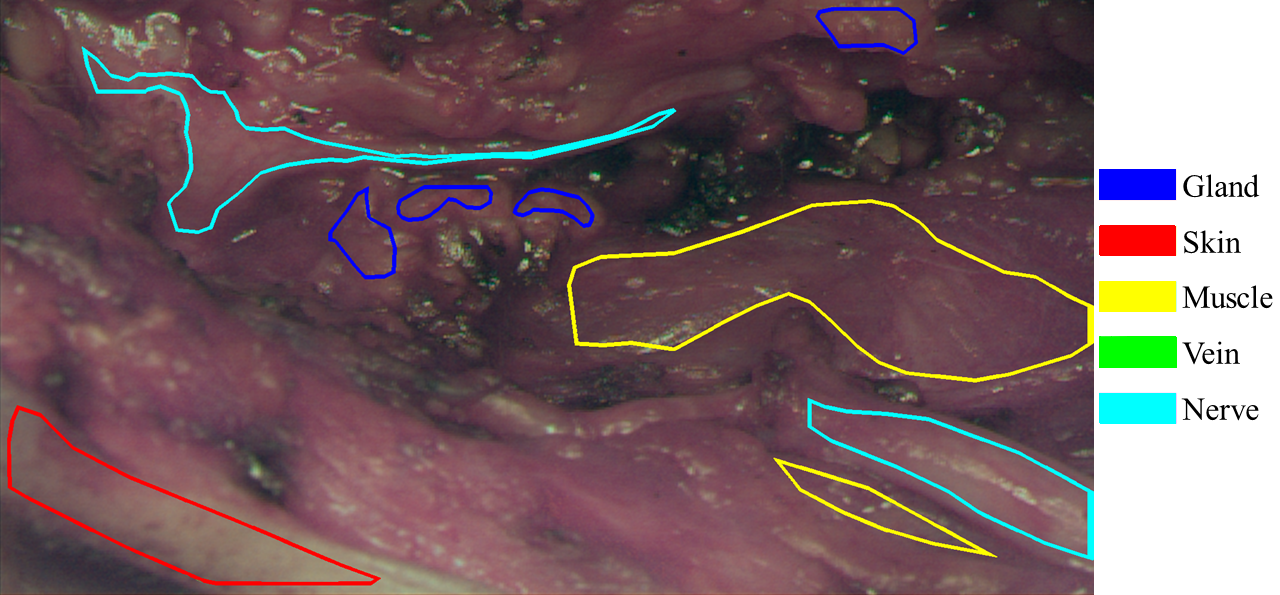}
	\caption{The different tissue regions, annotated by the surgeon, are bordered using specific color coding.}
	\label{img:annotate}
\end{figure}
\begin{figure}[t] \centering
	\includegraphics[width=.55\columnwidth]{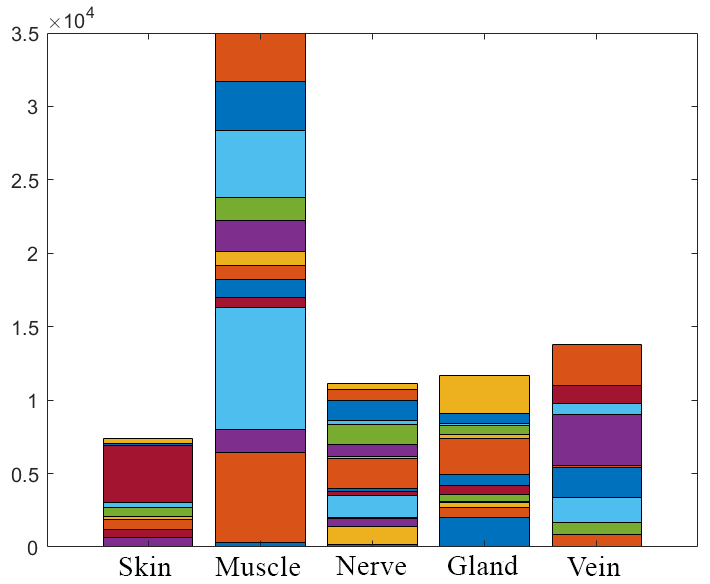}
	\caption{The distribution of all recorded tissue patches according to their categorization into the five tissue classes.}
	\label{img:data}
\end{figure}

The annotations of the individual tissue types were carried out by the performing surgeon directly after image acquisition, resulting in specific well defined tissue areas. Figure \ref{img:annotate} exemplary shows the annotation regions of Figure \ref{img:rgb}. The annotated tissue areas are divided into patches of $31\times31$ pixels in spatial dimension by a sliding window approach with stride of $10$ for training the network. Care is taken to ensure that only one type of tissue is included per patch. In total, more than $80.000$ patches has been created; however, the distribution of the patches across the individual tissue types is highly inhomogeneous, with more than $43\%$ of the patches being muscle, cf.~Figure \ref{img:data}. 

For evaluation, we followed the principle `leave one subject out' and separated all tissue data of specific patients from the remaining training data. It was ensured that patients were selected in such a way that at least $12\%$ of each tissue type are included in the evaluation set. As a result, three patients are excluded from the training and used as a test set. The remaining patches are used for training, with the data randomly divided into a training group ($92\%$) and a validation group ($8\%$).
For training the network, we used maximum-likelihood with Adam optimizer, a learning rate of $1.2\times10^{-4}$ and \textit{ReducedLROnPlateau} scheduler.

\section{Results and Discussion} \label{sec:res}
\begin{figure}[t] \centering
	\includegraphics[width=.7\columnwidth]{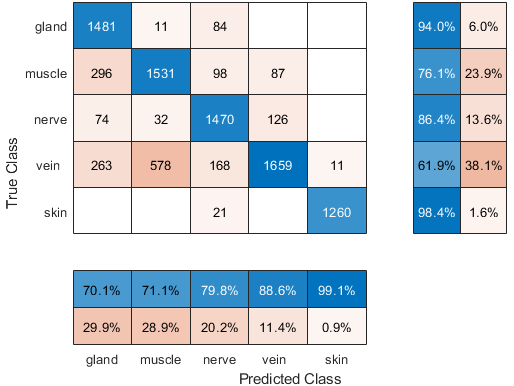}
	\caption{The confusion matrix of the best performing model applied on the evaluation patches showing an overall classification accuracy of $83.4\%$.}
	\label{img:conf}
\end{figure}

In the validation phase, the model demonstrated a commendable overall accuracy of $98.7\%$, indicating its robustness and effectiveness following thorough training despite the unbalanced data. 
Subsequent evaluation on patches from excluded patients yielded an overall accuracy of $83.4\%$, underscoring the model's proficiency in detecting and identifying tissue types. Individual sensitivity and specificity metrics, as depicted in Figure \ref{img:conf}, further elucidate the model's performance across different tissue categories. Notably, skin exhibited outstanding sensitivity and specificity, both in the range of $99\%$, highlighting the model's aptitude in discerning this tissue type. Glandular tissue followed suit with a notable sensitivity of $94\%$. Moreover, the model displayed robust performance in identifying high-risk tissue like nerves, with a sensitivity exceeding $86\%$. However, vein detection proved to be a more challenging task, exhibiting a lower sensitivity of approximately $62\%$. Nevertheless, specificity remained consistently high across all tissue types, surpassing $70\%$. 

These patch-wise analysis unequivocally demonstrate the feasibility of achieving reliable intraoperative tissue differentiation using our hyperspectral imaging system. Particularly noteworthy is the model's high sensitivity in detecting parotid or nerve tissue, underscoring its clinical significance in preserving and differentiating critical anatomical structures during parotidectomy. Furthermore, the observation that veins were frequently confused with muscle underscores potential areas for refinement and optimization in future iterations of the model. This highlights the need for additional training data, which should be more meticulously balanced across all tissue types, to enhance the model's ability to distinguish between different tissue classes accurately.

\begin{figure}[t]
	\includegraphics[width=\columnwidth]{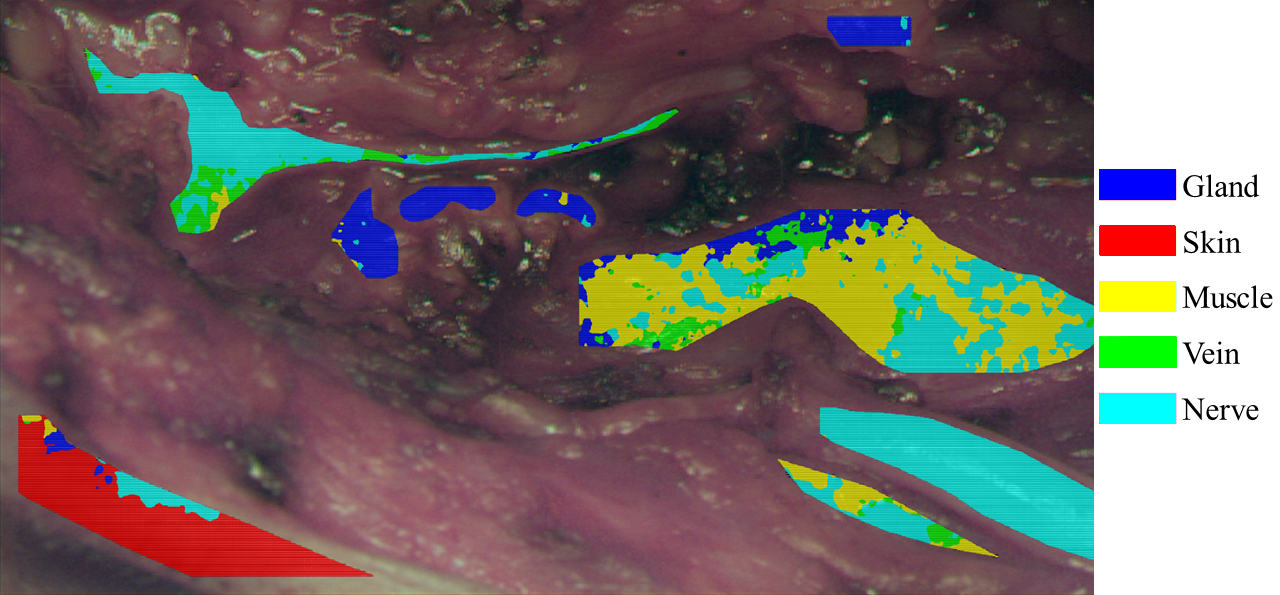}
	\caption{Predicted areas in an evaluation image, cf.~Fig.~\ref{img:annotate}.}
	\label{img:result}
\end{figure}

If the trained network is not applied to individual patches but to complete hyperspectral images, a visualization of the tissue classification can be obtained by overlaying the prediction results on the calculated RGB representation. Such augmented reality (AR) visualization holds potential for intraoperative use, providing surgeons with real-time feedback on tissue classification. A sample output of the best-trained CNN is visualized in Figure \ref{img:result}. For analytical purposes, only previously annotated tissue areas are visualized with the network's prediction. Across all classes, the visualization demonstrates good recognition performance, mirroring the same overall accuracy as the patch-wise analysis. Notably, muscle exhibits the poorest prediction, consistent with the findings of the patch-wise analysis as depicted in Figure \ref{img:conf}. Unfortunately, veins are not visible in this example, i.e., they have not been annotated. Nevertheless, individual pixels in other tissue areas have been classified as veins.

Upon closer examination of the pixel-accurate visualization of the annotated regions in the excluded patients, it becomes evident that individual tissue areas exhibit greater fluctuation. Small regions of incorrectly predicted tissue emerge within larger correctly determined tissue areas. This phenomenon is attributed to locally occurring specular highlights of the surgical illumination on the moist tissue surface. These specular highlights totally reflect the light source and hold no information about the reflecting object, i.e., the present tissue type. As the sensor saturates in these areas, robust classification becomes unattainable. In a further post-processing step, this behavior could be remedied by incorporating the classification results already obtained from neighboring areas. By leveraging the contextual information provided by nearby pixels, the model can better adapt to variations caused by localized reflections, thus improving the overall robustness of tissue classification. This approach allows for a more nuanced interpretation of the hyperspectral data, mitigating the impact of sensor saturation and enhancing the accuracy of tissue classification in challenging regions.

\section{Conclusion} \label{sec:conc}
In the present study, despite the constraints of a relatively small sample size, our utilization of a stereo-HSI system in conjunction with an CNN has facilitated the accurate identification of five distinct tissue classes crucial for parotidectomy. Our findings underscore the feasibility and efficacy of employing HSI technology in intraoperative settings, where real-time tissue differentiation is paramount. Moreover, the utilization of intraoperatively acquired human data further enhances the clinical relevance and applicability of our approach. Importantly, our achieved results stand in line with those reported in prior research, e.g., by Barberio et al.~\cite{BarberioDL}, thus validating the robustness and generalizability of our methodology.
Future investigations will aim to expand the amount of training data, the scope of tissue classes, optimize model performance and introduce post-processing techniques to best meet the challenges in surgical scenarios.


\textsf{\textbf{Author Statement}} This work was funded by the German BMBF under Grant No.~16SV8602 (KIPos). The authors state no conflict of interest. Informed consent has been obtained from all individuals included in this study. The research related to human use complies with all the relevant national regulations, institutional policies and was performed in accordance with the tenets of the Helsinki Declaration, and has been approved by the Charit\'e ethics board.


\begin{thebibliography}{99}


\bibitem{BarberioDL}
Barberio M, Collins T, Bencteux V, Nkusi R, Felli E, Viola MG, Marescaux J, Hostettler A, Diana M. Deep Learning Analysis of In Vivo Hyperspectral Images for Automated Intraoperative Nerve Detection. Diagnostics, 2021, 11:1508.

\bibitem{ClancySSI}
Clancy NT, Jones G, Maier-Hein L, Elson DS, Stoyanov D. Surgical spectral imaging. Med Image Anal, 2020, 63:101699.

\bibitem{EbnerSnapshot}
Ebner M, Nabavi E, Shapey J, Xie Y, Liebmann F, Spirig JM, et al. Intraoperative hyperspectral label-free imaging: from system design to first-in-patient translation. J Phys D: Appl Phys, 2021, 54(29):294003.

\bibitem{LuHSIreview}
Lu G, Fei B. Medical hyperspectral imaging: a review. J Biomed Opt, 2014, 19(1):010901.

\bibitem{MuhleComparison}
Mühle R, Markgraf W, Hilsmann A, Malberg H, Eisert P, Wisotzky EL. Comparison of different spectral cameras for image-guided organ transplantation. J Biomed Opt, 2021, 26(7):076007.

\bibitem{Ortac21}
Ortac G, Giyasettin O. Comparative study of hyperspectral image classification by multidimensional Convolutional Neural Network approaches to improve accuracy. Expert Systems with Applications, 2021, 182:115280.

\bibitem{ShapeyReview}
Shapey J, Xie Y, Nabavi E, Bradford R, Saeed SR, Ourselin S, Vercauteren T. Intraoperative multispectral and hyperspectral label-free imaging: A systematic review of in vivo clinical studies. J Biophotonics, 2019, 12(9):e201800455.

\bibitem{FischerOrgan}
Studier-Fischer A, Seidlitz S, Sellner J, Özdemir B et al. Spectral organ fingerprints for machine learning-based intraoperative tissue classification with hyperspectral imaging in a porcine model. Scientific Reports, 2022, 12(1):11028.

\bibitem{WisotzkyDemosaicking}
Wisotzky EL, Daudkane C, Hilsmann A, Eisert P. Hyperspectral Demosaicing of Snapshot Camera Images Using Deep Learning. Pattern Recognition: 44th DAGM German Conference. DAGM GCPR, 2022, pp. 198-212.

\bibitem{WisotzkyValidation}
Wisotzky EL, Kossack B, Uecker FC, Arens P et al. Validation of two techniques for intraoperative hyperspectral human tissue determination. J Med Imaging, 2020, 7(6):065001.

\bibitem{WisotzkySensor}
Wisotzky EL, Rosenthal JC, Wege U, Hilsmann A, Eisert P, Uecker FC. Surgical guidance for removal of cholesteatoma using a multispectral 3D-endoscope. Sensors, 2020, 20(18):5334.

\bibitem{WisotzkyStereoHSI2}
Wisotzky EL, Triller J, Hilsmann A, Eisert P. Multispectral Stereo-Image Fusion for 3D Hyperspectral Scene Reconstruction. Proceedings of the 19th International VISIGRAPP, 2024, pp. 88-99.

\bibitem{WisotzkyStereoHSI}
Wisotzky EL, Triller J, Kossack B, Globke B et al. From Multispectral-Stereo to Intraoperative Hyperspectral Imaging: a Feasibility Study. Curr Dir Biomed Eng, 2023, 9(1):311-314

\bibitem{WisotzkyDemosaicking2}
Wisotzky EL, Wallburg L, Hilsmann A, Eisert P, Wittenberg T, G\"ob Stephan. Efficient and Accurate Hyperspectral Image Demosaicing with Neural Network Architectures. Proceedings of the 19th International VISIGRAPP, 2024, pp. 541-550.

\end{thebibliography}

\end{document}